\documentclass[draft,jgrga]{agutex}

\usepackage{lineno}
\usepackage[pdftex]{graphicx}

\setkeys{Gin}{draft=false}
\authorrunninghead{LUKIANOVA ET AL.}

\titlerunninghead{MONTHLY SOLAR WIND SPEEDS}

\begin{document}

%
%

\title{Centennial evolution of monthly solar wind speeds: Fastest monthly solar wind speeds from long-duration coronal holes}

%

%
%


\authors{R. Lukianova\altaffilmark{1,2,3}, L. Holappa\altaffilmark{1} and K. Mursula\altaffilmark{1}} 
\altaffiltext{1}{ReSoLVE Centre of Excellence, Space Physics Research Unit, University of Oulu, Oulu, Finland}
\altaffiltext{2}{Geophysical Center of Russian Academy of Science, Moscow, Russia}
\altaffiltext{3}{Space Research Institute, Moscow, Russia}







%
%


\begin{abstract}

High speed solar wind streams (HSSs) are very efficient drivers of geomagnetic activity at high latitudes. 
In this paper we use a recently developed $\Delta{H}$ parameter of geomagnetic activity, calculated from the night-side hourly magnetic field measurements of the Sodankyl\"a observatory, as a proxy for solar wind (SW) speed at monthly time resolution in 1914-2014 (solar cycles 15-24). 
The seasonal variation in the relation between monthly $\Delta{H}$ and solar wind speed is taken into account by calculating separate regressions between $\Delta{H}$ and SW speed for each month.
Thereby, we obtain a homogeneous series of proxy values for monthly solar wind speed for the last 100 years.
We find that the strongest HSS-active months of each solar cycle occur in the declining phase, in years 1919, 1930, 1941, 1952, 1959, 1973, 1982, 1994 and 2003.
Practically all these years are the same or adjacent to the years of annual maximum solar wind speeds.
This implies that the most persistent coronal holes, lasting for several solar rotations and leading to the highest annual SW speeds, are also the sources of the highest monthly SW speeds.
Accordingly, during the last 100 years, there were no coronal holes of short duration (of about one solar rotation) that would produce faster monthly (or solar rotation) averaged solar wind than the most long-living coronal holes in each solar cycle produce.

\end{abstract}

%
%

%

\begin{article}

%
%

\section{Introduction}

Solar wind (SW) speed is one of the most important factors in the solar wind-magnetosphere interaction. 
Long-term averages of SW speed are strongly modulated by the occurrence of high-speed streams (HSSs), which are known to originate from coronal holes \citep{Krieger_1973, Gosling_1976, Kojima_1990}.
The occurrence of HSSs at the Earth's orbit maximizes during the declining phase of the solar cycle, when high speed streams from equatorward extensions of polar coronal holes often reach low heliographic latitudes and the ecliptic plane \citep{Hakamada_1981}. 
HSSs have a strong effect on geomagnetic activity especially at high latitudes and the auroral zone. 
For example, the occurrence of magnetospheric substorms is strongly modulated by the occurrence of HSSs [\citeauthor{Tanskanen_2005}, \citeyear{Tanskanen_2005}, \citeyear{Tanskanen_2011}]. Recently, \citet{Lukianova_2012} showed that magnetic disturbances caused by HSSs can be seen even in annual means of the vertical magnetic field component (Z) on the polar cap and the horizontal magnetic field component (H) at auroral latitudes. 
HSSs include enhanced Alfv\'en wave activity, which lead to a repeated occurrence of substorms \citep{Lyons_2009} and energetic particles \citep{Denton_2012}.
The HSS-related magnetic disturbances are mainly reflected in the westward auroral electrojet (WEJ), which is enhanced during substorms.
\citet{Lukianova_2012} found that the highest intensity of WEJ occurred in 2003, in the declining phase of the solar cycle (SC) 23, leading to a major reduction of H at auroral latitudes and strengthening of Z within the northern and southern polar caps. 

\citet{Mursula_2015} exploited the strong correlation between the high-latitude magnetic disturbances and SW speed and reconstructed the annual means of SW speed from two magnetic stations (Godhavn/Qeqertarsuaq, GDH and Sodankyl\"a, SOD) with the longest series of high-latitude geomagnetic data since 1926 (GDH) and 1914 (SOD). 
They thus covered most solar cycles of the so called Grand Modern Maximum (GMM) of solar activity, during which solar activity has been greater than centuries to millennia before \citep{Solanki_2000}. 
\citet{Mursula_2015} found that high annual speeds occurred in the declining phase of all SCs 16–-23. 
The estimated annual SW speed exceeded 500 km/s in 1930, 1941, 1951-1953, 1963, 1974, 1994 and 2003. 
In the early 1950s, high HSS activity continued for three successive years, with the highest yearly activity, up to 570 km/s, found in the year 1952. 
They noted that cycle 19, which marks the sunspot maximum period of the GMM, was preceded by exceptionally strong polar fields during the declining phase of cycle 18, which proves the $\Omega$-mechanism (conversion of poloidal fields to toroidal fields) of the solar dynamo theory \citep{Babcock_1961} for this period of very high solar activity. 
The aim of this paper is to study the long-term evolution HSSs at monthly time resolution, using the same methodology as previously applied by \citet{Lukianova_2012} and \citet{Mursula_2015}. 
This paper is organized as follows. In Section 2 we present the data and introduce the $\Delta{H}$ parameter.
In Section 3 we study the relation between monthly $\Delta{H}$ and SW speed.
In Section 4 we present monthly proxies of SW speed. 
Discussion and conclusions are given in Section 5.

\section{Data}

We use hourly measurements of horizontal magnetic field at the Sodankyl\"a geophysical observatory (SOD; geographic latitude and longitude: $67.37^{\circ}$, $26.63^{\circ}$; geomagnetic latitude and longitude: $64^{\circ}$, $119^{\circ}$), located near the equatorward boundary of the auroral oval. 
At Sodankyl\"a, recordings of the Earth's magnetic field vector have been made since 1914 (interrupted only during the World War II in 1945), forming the longest continuous high-latitude geomagnetic measurements available for more than 100 years. 

Because of the proximity of SOD to the auroral electrojets, the largest perturbations occur in the geomagnetic horizontal H component which is directed to the magnetic north. During magnetospheric substorms, WEJ increases during the growth and expansion phases and decays back to quiet-time level during the recovery phase \citep{Akasofu_1964}. 
The WEJ-related magnetic disturbances are mainly manifested in the midnight magnetic local time (MLT) sector and the eastward auroral electrojet in the late afternoon sector. 
Accordingly, the amplitude of the daily curve varies in the range of approximately -500 to 500 nT. 
At SOD, the 4-hour time interval of 20-23 UT (22-01 LT) is the most appropriate time to estimate the WEJ intensity.

We define the geomagnetic disturbance parameter $\Delta{H}$ as follows. 
For each month we calculate the quiet-time level H(q) by averaging the night-time (20-23 UT) H values during the five quietest days. The quietest days have been calculated from the local K indices and they are available at the SGO database. 
Then we calculate $\Delta{H}$ by calculating the difference between H(q) and the monthly average of all night-time (20-23 UT) H values. 
($\Delta{H}$ = H(q) - H is positive since WEJ reduces H).

We also use the measured hourly SW speed (V) values of the OMNI data base (\texttt{http://omniweb.gsfc.nasa.gov/}) since 1964. To quantify the relationship between $\Delta{H}$ and SW speed we calculate a linear regression separately for different months. Because of numerous data gaps in the OMNI data base especially during 1980s and early 1990s, monthly means of SW speed cannot always be reliably calculated. 
In the regressions we only use those values of $\Delta{H}$ and SW speed when both parameters have been measured simultaneously. Moreover, in order to have sufficient statistics for each month we neglect, when calculating the regressions, those months when the data coverage is less than 30\%.
Overall, there are 55 months (i.e., 9\%) neglected by this requirement.
The problem is worst in the early 1980s, when up to 8 months were neglected in the year 1984.

\section{Relationship between monthly solar wind speed and $\Delta{H}$}

Figure 1 shows, as an example, the correlation between monthly $\Delta{H}$ and SW speed for Januaries in 1964-2014.
The correlation is fairly linear and statistically significant (correlation coefficient is 0.76; zero correlation probability p = 0.0002 using a first order autoregressive (AR-1) noise model), and there are no actual outliers in Fig. 1.
Note that one station cannot always be exactly at the site of the maximum WEJ enhancement related to the substorm current wedge, which produces a source of error in Fig. 1. 
Also the CMEs contribute to increasing scatter, especially in solar maximum years (see later).
Data gaps in SW measurements (to be discussed in detail in Section 4) are a source of significant scatter in Fig. 1.
Also one can see from Figure 1 that data points lie symmetrically around the regression line for the whole range of $\Delta{H}$. 
This is better seen in the bottom panel of Fig. 1 which depicts the residuals of the regression, i.e., the differences between the measured and estimated monthly SW speeds.
The homoscedasticity of the residuals guarantees that the standard least squares fit performs well and can be reliably used to reconstruct the monthly means of SW speed. Table 1 gives the regression coefficients and correlation coefficients for similar fits for each month. 

Figure 2 depicts the regression coefficients (slope and intercept) for each month. Figure 2 shows that while the intercept varies very little from month to month, the relative variation in the slope is much larger. The slope maximizes during mid-winter (Dec and Jan) and mid-summer (Jun), and minimizes around equinox months. Thus, the monthly average SW speed required to produce a given value of $\Delta{H}$ is largest during mid-winter and mid-summer, and the response at high latitudes ($\Delta{H}$) to SW speed is stronger (slope is lower) during equinoxes. 
This is most likely related to the semiannual variation of geomagnetic activity, whose main drivers are the equinoctial mechanism \citep{Cliver_2000, Lyatsky_2001} related to the seasonally changing ionospheric conductivity and the Russell-McPherron mechanism \citep{Russell_1973, McPherron_2009}, where the solar equatorial magnetic field gets projected to the southward component in the GSM coordinate system. 
Here we use the monthly regressions and related parameters to reconstruct the monthly SW speeds in the early 20th century.
 
Figure 3a shows the scatter plot of the actually measured SW speeds and estimated SW speeds (using the regression parameters in Table 1) for all months in 1964-2014. 
(Here no requirement for data coverage is imposed). 
One can see that the relation between the estimated and measured values is fairly linear and there are only two large outliers in the fit. (These two outliers, June and Feb 1982, were neglected when calculating the regressions). 
Figure 3b shows the fit residuals $\delta$ = $V(measured)$ - $V(estimated)$ as a function of estimated SW speeds. 
One can see that the majority of the residuals over almost the whole range of values are homoscedastically distributed. 
Note that the data gaps in the solar wind measurements and from the non-HSS related SW effects in $\Delta{H}$ (mainly due to coronal mass ejections, CMEs) increase scatter and may increase the number of outliers. 
We will discuss the data gaps in more detail in Section 4.

\section{Monthly solar wind speeds}

The above described monthly regressions between SW speed and $\Delta{H}$ are applied to reconstruct the monthly SW speeds during the last 100 years since 1914. Figure 4 depicts the monthly SW speed proxies estimated from $\Delta{H}$ in 1914-2014 together with their $\pm 1\sigma$ errors, and the measured monthly SW speeds in 1964-2014, separately for each month. (We have included in Figure 4 all available data for both the measured and proxy values). Figure 4 shows that, even at the monthly timescale, the $\Delta{H}$ based proxies represent the SW speed with reasonable accuracy. 
The proxy covers the range of the measured SW speeds relatively well, except for some of the highest peaks and the lowest minima. There are some differences especially from 1980s until mid-l990s when there are numerous data gaps in the SW measurements, and the lower statistics increases random fluctuations.

Figure 5 shows the proxy and the measured monthly SW speeds in 1964-2014 for months with $30$\% data coverage condition imposed. Therefore Figure 5 includes more data gaps than Figure 4. 
Here we have included only those hours when the two parameters had simultaneous measurements. 
The bottom panel of Figure 5 depicts the monthly fractions of data gaps in SW measurements. 
The data gaps practically end in 1995, since when the SW is continuously measured by the ACE (and later Wind) satellite. At this time the accuracy of the proxy values is also somewhat improved. The standard deviation of the difference between the estimated and measured values of SW speed is 31 km/s in 1995-2014 and 39 km/s in 1964-1994. 

Despite the differences between the measured and the proxy values of monthly SW speeds, the highest peaks of each solar cycle tend to agree with each other. This is the case for solar cycle 20 (when the measured and the proxy peaks occur in April and March 1973), for cycle 22 (both in February 1994) and cycle 23 (both in June 2003). This comparison fails only for solar cycle 21, when the two peaks are in different years (April 1983 and Feb 1982). The largest number of data gaps take place in the early 1980s, in the declining phase of SC 21, when IMP-8 satellite was the only satellite measuring the solar wind. 
However, we suspect that the difference between the measured and the proxy values especially during the high peaks in 1982 is not only due to the data gaps. These peaks occurred close to the maximum of solar cycle 21 and are most likely significantly affected by CMEs.
Note that CMEs tend to produce exceptionally high values for the proxy due to its other geo-effective factors. This is seen clearly in 1982--1983.

Figure 6 shows the reconstructed monthly SW speeds in 1914-2014 from Figure 4 as a single time series. Figure 6 shows that the highest peaks for each SC always occur during the declining phase of the solar cycle. The cycle peaks of monthly SW speeds of cycles 15-23 occurred in 1919, 1930, 1941, 1952, 1959, 1973, 1982, 1994 and 2003. These years are almost the same as the cycle peak years of the annual SW speeds from SOD (1918, 1930, 1941, 1952, 1959, 1974, 1984, 1994 and 2003; almost the same years were found in the GDH station as well) estimated by Mursula et al. (2015). 
So, in 6 out of 9 cycles the cycle peaks for monthly and yearly peaks in the SW speed proxy occur in the same year. In two cycles they are found in successive years, when the high-speed stream forming the monthly maximum is very likely produced by the coronal hole which yields the annual speed maximum in the adjacent year.
This is the case for 1918-1919, when the monthly peak is found in May 1919 (520 km/s) while almost equally high monthly values were found already in 1918 (e.g., in December 520 km/s), which was the year of the most persistent yearly HSS activity. 
Only for cycle 21, when three years 1982-1984 all have roughly equal annual SW speed values, the respective peaks have a difference of 2 years. 

The fact that practically all of the highest monthly speeds are found during the same or adjacent years as the highest annual speeds implies that the most persistent coronal holes, which are responsible for the highest annual means of SW speed, are also the sources of the highest monthly SW speeds. 
Persistent coronal holes can live for several months (up to one year), whence the highest monthly SW speed value in one calendar year can, actually, be produced by a persistent coronal hole extending to or from the adjacent year, where it forms the annual maximum occurrence of high-speed streams (maximum of annual solar wind speed). 
These results imply that it is very unlikely that highest speed streams of roughly one-month (or solar rotation) duration would appear outside of the times when the most persistent coronal holes appear in the Sun.
These uniform results also strongly support the method used here to estimate the monthly SW speeds from the $\Delta{H}$ parameter.

\section{Discussion and conclusions}

In this paper we have utilized the longest available high-latitude measurement of the geomagnetic field made at Sodankyl\"a, Finland, and used a local night-time measure ($\Delta{H}$) of geomagnetic activity to estimate the strength of the westward auroral electrojet, which is a sensitive proxy of SW speed. 
We have calculated linear regressions between $\Delta{H}$ and SW speed separately for all months. 
Even at monthly timescale we find high correlations between the two parameters for all months, giving evidence that other factors in solar wind, especially the intensity of the interplanetary magnetic fields which is enhanced during CMEs [Richardson and Cane, 2012], have a significantly smaller effect for the monthly averages of $\Delta{H}$ at Sodankyl\"a. This supports the earlier studies which have shown the importance of the SW speed for substorm occurrence \citep{Tanskanen_2005} and high-latitude geomagnetic activity \citep{Finch_2008, Lukianova_2012, Holappa_2014}.

The relation between $\Delta{H}$ and SW speed shows a clear seasonal variation so that during equinoxes the coupling is stronger, i.e., a given SW speed value yields a higher value of $\Delta{H}$. This seasonal variation is most likely related to the semiannual variation of geomagnetic activity, mainly due to the equinoctial \citep{Cliver_2000} and Russell-McPherron effects \citep{Russell_1973, McPherron_2009}, which modulate the geoeffectiveness of HSSs. We take this seasonal variation into account by determining regression coefficients separately for different months. 

Using the monthly regressions we have estimated the monthly means of the SW speed for the last 100 years (1914-2014). We find that the largest monthly SW speeds, i.e., the highest HSS-active months in each solar cycle occur in the declining phase of the cycle, in the years 1919, 1930, 1941, 1952, 1959, 1973, 1982, 1994 and 2003 for cycles 15-23, respectively. 
This confirms the observation based on annual means that the most persistent high-speed streams occur during the declining phase of all cycles during the last century \citep{Mursula_2015}, and extends this observation to the shorter-living streams with duration of about one solar rotation.

Interestingly, for 8 out of 9 solar cycles studied (all except for cycle 21 when the statistics of SW measurement  in 1980s was poor), the years with the highest monthly SW speeds of the respective cycle are the same or adjacent years to the peak years based on the annual SW speeds. This suggests that the most persistent coronal holes lasting for several months (solar rotations) are also the sources of the highest monthly (one solar rotation) SW speeds. Accordingly, no short-term coronal holes are found that would be large or effective enough to produce the highest monthly SW speed of any solar cycle. 

In seven months (May-June 1930, February-March 1952, April-May 1994 and June 2003) the monthly mean SW speed based on the $\Delta{H}$ proxy exceeded 550 km/s. 
All these months occur in years of the highest annual solar wind speed in the respective cycle (16, 18, 22 and 23).
This further supports the idea that coronal holes emitting fast SW speed indeed tend to be persistent and live longer than one solar rotation. 

The temporal distribution of the highest SW speeds is interesting. We find strong HSS activity during the rather weak sunspot cycle 16, which compares with the HSS activity of the later, more active sunspot cycles. 
Thus, although the mean level of solar wind speed slightly changes (increases) with the long-term evolution of sunspot activity, the occurrence of the highest SW speeds (i.e., coronal holes) does not follow them very closely.


%

%
%
%
%
%

%
%
%
%

\begin{acknowledgments}
We acknowledge the financial support by the Academy of Finland to the ReSoLVE Centre of Excellence (project no. 272157).
We thank the Sodankyl\"a Geophysical Observatory for providing the magnetic field data at (\texttt{http://www.sgo.fi/}).
The solar wind data were downloaded from the OMNI2 database (\texttt{http://omniweb.gsfc.nasa.gov/}).
\end{acknowledgments}

%
%
%
%
%
%
%
%
%
%






%
%

\end{article}



%
%
%
%
%
%

\newpage


\begin{table}
\begin{center}
\begin{tabular}{c|cccc}\hline
Month & a [km/s/nT] & b [km/s] & cc & p \\\hline
1 & 1.87 & 367 & 0.71 & $2\cdot 10^{-5}$ \\
2 & 1.62 & 363 & 0.74 & $7\cdot 10^{-7}$ \\
3 & 0.99 & 377 & 0.72 & $4\cdot 10^{-6}$ \\
4 & 0.97 & 373 & 0.77 & $2\cdot 10^{-7}$ \\
5 & 1.26 & 359 & 0.82 & $7\cdot 10^{-9}$ \\
6 & 1.69 & 360 & 0.73 & $3\cdot 10^{-7}$ \\
7 & 1.34 & 372 & 0.74 & $2\cdot 10^{-4}$ \\
8 & 1.00 & 386 & 0.64 & $4\cdot 10^{-6}$ \\
9 & 0.85 & 373 & 0.68 & $8\cdot 10^{-6}$ \\
10& 0.72 & 379 & 0.54 & $1\cdot 10^{-2}$ \\
11& 1.05 & 374 & 0.73 & $5\cdot 10^{-4}$ \\
12& 1.79 & 363 & 0.77 & $2\cdot 10^{-5}$ \\\hline
\end{tabular}\caption{Slope (a), intercept (b), correlation coefficient (cc) and p-values of the regression line $V = a\cdot \Delta{H} + b$ for each month.}
\label{table}
\end{center}
\end{table}


\newpage

\clearpage

\begin{figure}
\includegraphics[width=\linewidth]{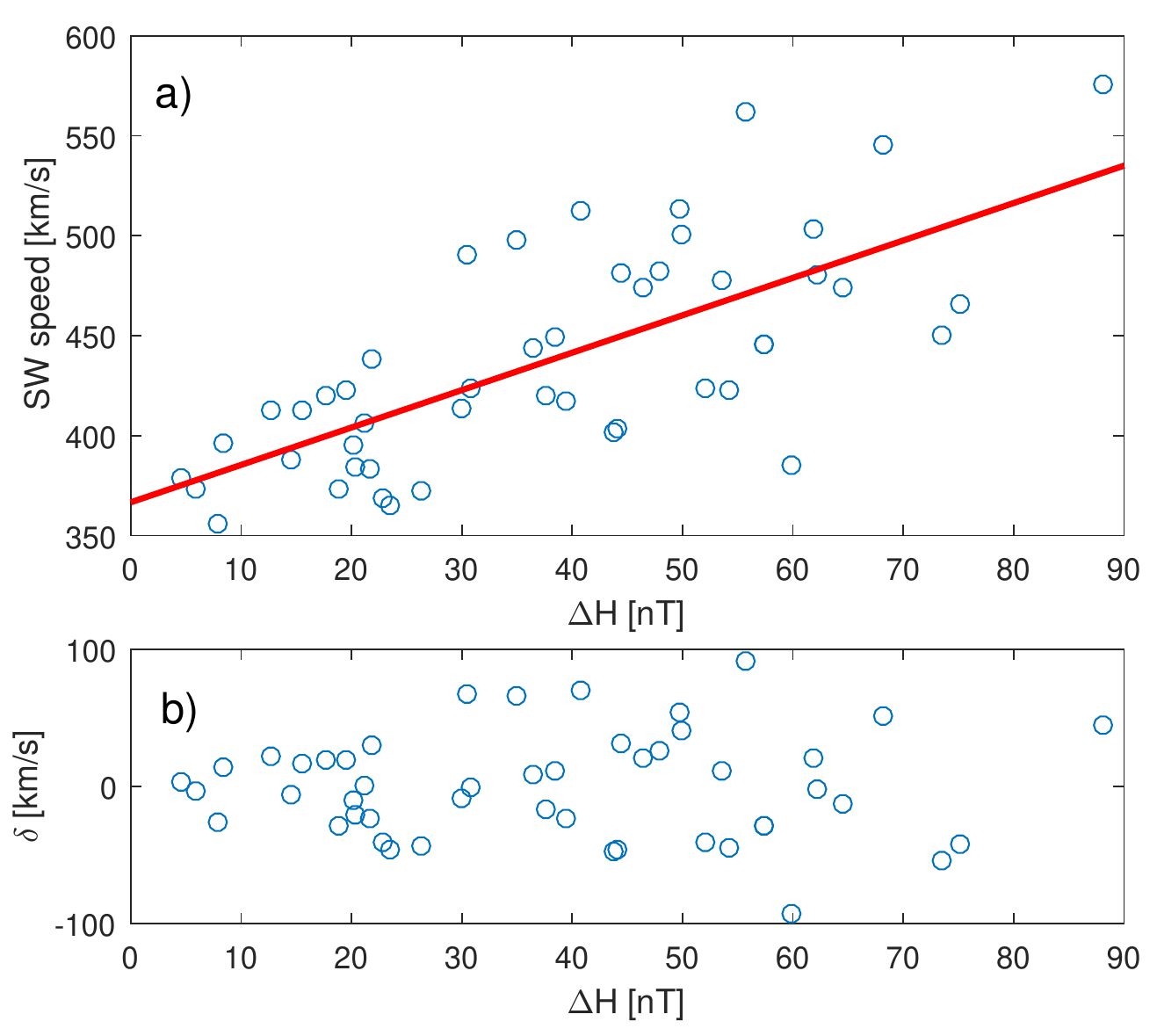}
\caption{a) Scatter plot of the monthly values of SW speed and $\Delta{H}$ observed in Januaries 1964-2014. The red line shows the regression line. b) Differences of the measured and estimated values of SW speed as a function of the estimated speed values.}
\label{fig1}
\end{figure}


\begin{figure}
\includegraphics[width=\linewidth]{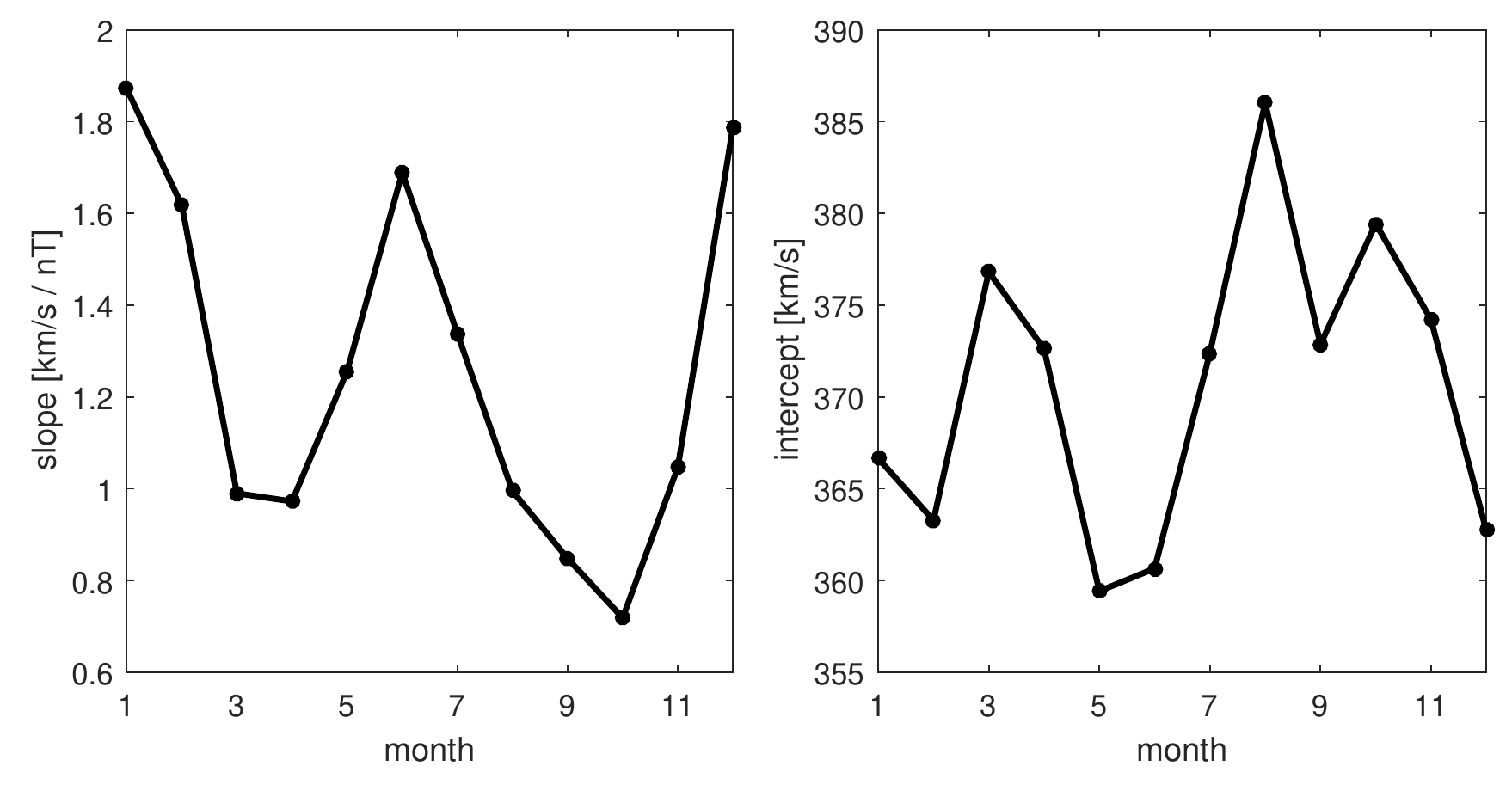}
\caption{Slope (a) and intercept (b) of the regression line $V = a\cdot \Delta{H} + b$ for each month.}
\label{fig2}
\end{figure}


\begin{figure}
\includegraphics[width=\linewidth]{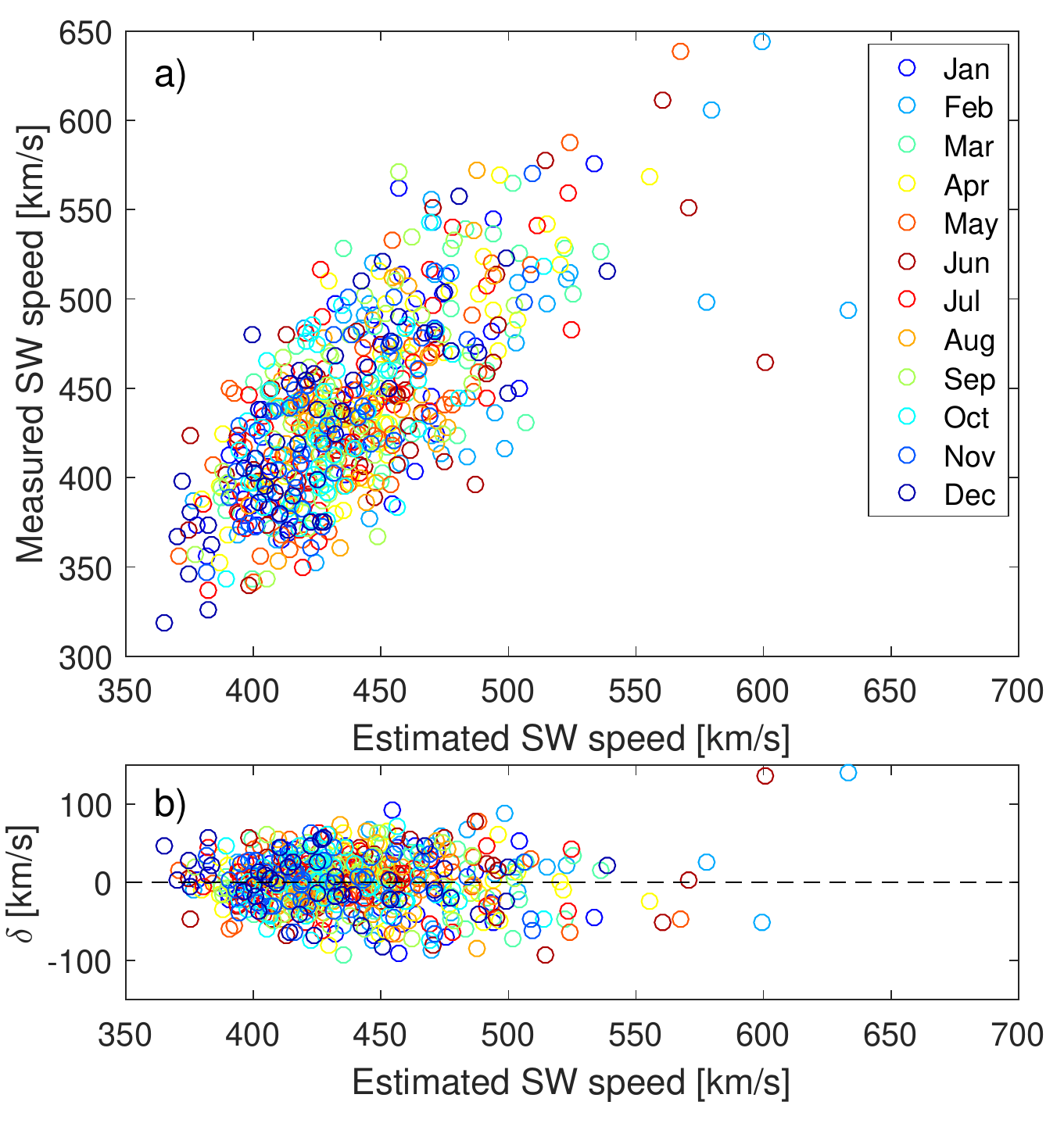}
\caption{a) Scatter plot of estimated and measured monthly SW speeds for all months in 1964-2014. Colors indicate the respective month. b) Differences of the measured and estimated values of SW speed as a function of the estimated speed values.}
\label{fig3}
\end{figure}


\begin{figure}
\includegraphics[width=\linewidth]{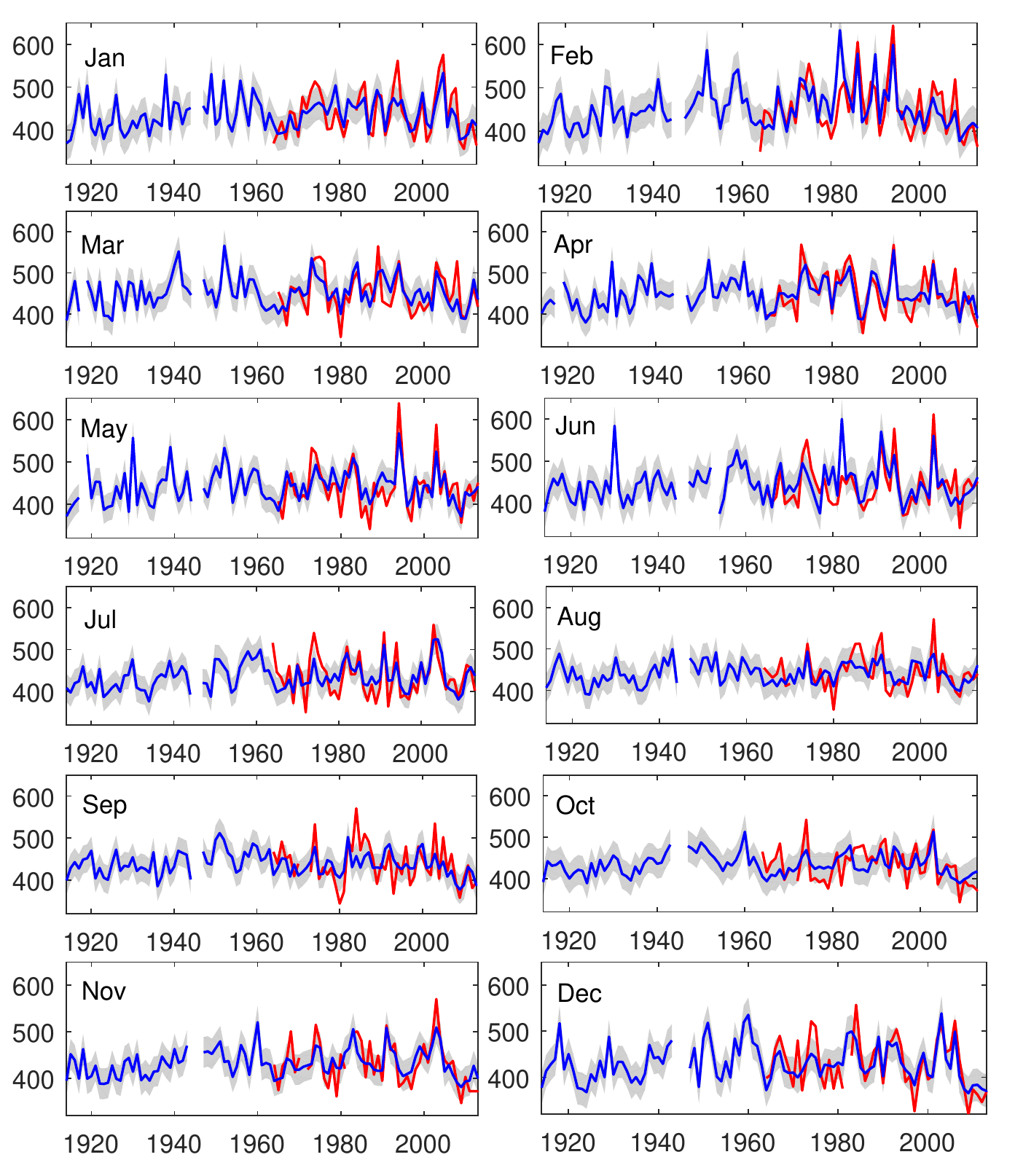}
\caption{a) Monthly SW speeds estimated from $\Delta{H}$ parameter (blue line) together with $\pm 1\sigma$ errors (shaded gray area) for separate months in 1914-2014, and the measured monthly SW speeds (red line).}
\label{fig4}
\end{figure}


\begin{figure}
\includegraphics[width=\linewidth]{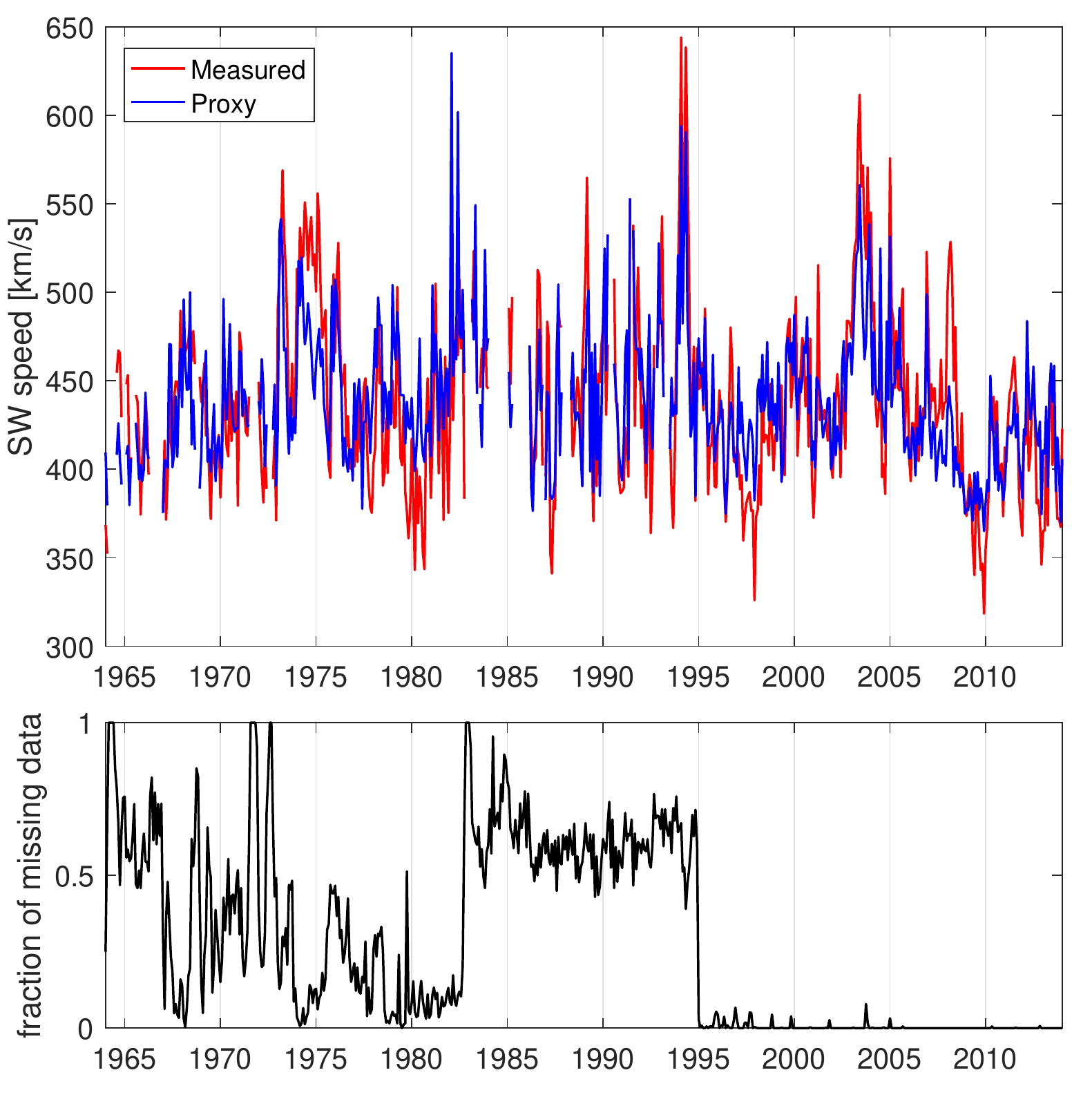}
\caption{Upper panel: Measured and estimated monthly SW speeds in 1964-2014. Only months with data coverage $>30$\% are shown. Lower panel: Monthly fraction of data gaps in SW measurements.}
\label{fig5}
\end{figure}


\begin{figure}
\includegraphics[width=\linewidth]{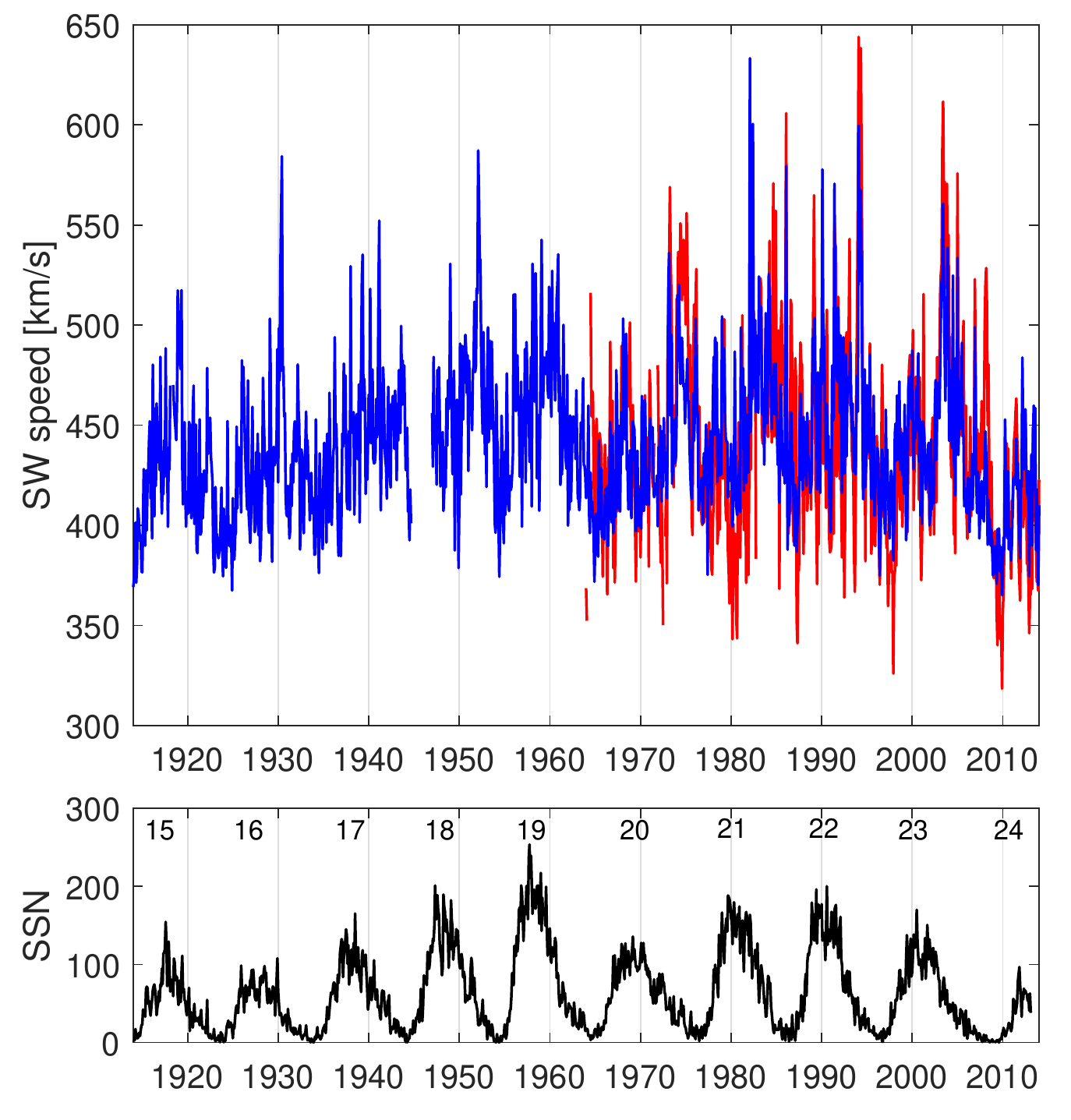}
\caption{Upper panel: Monthly averages of SW speed proxy estimated from the $\Delta{H}$ in 1914-1963 (blue line), the observed SW speed in 1964-2013 (red line). Bottom panel: monthly sunspot number and the solar cycle numbers.}
\label{fig6}
\end{figure}

\end{document}